# A Habitable Planet around HD 85512?

L. Kaltenegger[1,2], S. Udry[3] and F. Pepe[3]
[1]MPIA, Koenigstuhl 17, 69117 Heidelberg, Germany
[2]Harvard Smithsonian Center for Astrophysics, 60 Garden St., 02138 MA, Cambridge, USA, email: kaltenegger@mpia.de
[3]Observatoire de Geneve, Universit´e de Geneve, 51 ch. des Maillettes, CH–1290 Sauverny, Switzerland

**abstract**
*Aims*: In this study we assess the habitability of HD85512b, a 3.6$M_{Earth}$ planet orbiting a K5V star. The radial velocity data and orbital parameters for HD 85512 b have just been published, based on data from the dedicated HARPS-upgrade GTO program.
*Methods*: This paper outlines a simple approach to evaluate habitability of rocky planets from radial velocity (RV) searches by using atmospheric models of rocky planets with $H_2O/CO_2/N_2$ atmospheres, like Earth. We focus our analysis on HD 85512 b. To first order the limits of the Habitable Zone depend on the effective stellar flux distribution in wavelength and time, the planet's Bond albedo, and greenhouse gas effects in this approach. We also discuss the dependence of habitability on the measurement accuracies.
*Results*: We provide a simple set of parameters which can be used for evaluating current and future planet candidates from RV searches for their potential habitability. We find that HD 85512 b could be potentially habitable if the planet exhibits more than 50% cloud coverage. HD 85512 b is, with Gl 581d, the best candidate for exploring habitability to date, a planet on the edge of habitability.

## Introduction

The dedicated HARPS-Upgrade GTO program, recently published the radial-velocity data and the orbital parameters of a new low-mass planet of 3.6 ±0.5 $M_{Earth}$ minimum mass around the star HD 85512 (Pepe et al. 2011). HD 85512 is a K5V star with 0.126 ± 0.008 solar luminosity, 0.69 solar mass and an effecitve temperature, $T_{eff}$, of 4715 ± 102. HD 85512 b orbits its star in 58.43 ±0.13 days at 0.26 ± 0.005 AU with an eccentricity of 0.11 ± 0.1, which places it near the inner edge of the habitable zone (HZ). Here we assume that the planet's actual mass is close to its minimum mass. The stellar and planetary parameters for the system are given in table 1.

The planet is one of the least massive planets detected to date. It is the least massive planet confirmed in the Habitable Zone (HZ) of its host star. HD 85512 b provides a promising target for the study of habitability of exoplanets. Its mass is consistent with rocky planet models, an assumption we adopt for this paper.

Different aspects of what determines the boundaries of the HZ have been discussed broadly in the literature (see section 2). The width and distance of the HZ annulus - as well as the effective temperature - for an Earth-like atmosphere depends to a first approximation on 4 main parameters: 1) incident stellar flux which depends on stellar luminosity, spectral energy distribution and eccentricity of the system, 2) overall planetary Bond albedo, 3) greenhouse gas concentration, and 4) energy distribution in the planetary atmosphere. We compare these values to Earth, Venus as well as Gl 581c to assess the habitability of HD 85512 b.

The atmospheric model we present here only represents one possible nature of HD 85512 b – that of a rocky planet with an $H_2O/CO_2/N_2$ atmosphere, like Earth - in a wide parameter space (see results). This atmosphere model produces a dense water dominated atmosphere on the inner edge of the HZ and a dense $CO_2$ dominated atmosphere on the outer edge of the HZ. Between those limits, we assume that on a geological active planet, climate stability is provided by a feedback mechanism in which atmospheric $CO_2$ concentrations vary inversely with planetary surface temperature.

We introduce the concept of the Habitable Zone in section 2, explore the influence of first order effects and how they match the level of information and uncertainties for planetary candidates from radial velocity searches in section 3. In particular we focus on HD 85512 b. Section 4 states our conclusions.

Our approach can be applied to current and future candidates provided by RV searches.

**The concept of the Habitable Zone**

The habitable zone concept was first developed by Huang (1959, 1960) and has been calculated by several groups (see e.g. Kaltenegger & Sasselov 2011 and references therein, as well as Abbe et al. 2011 for limits of the HZ for dry planets). The main differences among these studies are the climatic constraints imposed. For this paper we focus on the circumstellar HZ (Kasting et al. 1993, Selsis et al. 2007), defined as an annulus around a star where a planet with a $CO_2/H_2O/N_2$ atmosphere and a sufficiently large water content like Earth can host liquid water permanently on a solid surface. Note that this definition of the HZ is adopted because it implies surface habitability and in turn allows remote detectability of life as we know it. Subsurface life that could exist on planets and moons with very different surface temperatures is not considered here, because of the lack of atmospheric features to remotely assert habitability (see e.g. Rosing 2005).

In this definition, the two edges of the HZ as well as the equilibrium temperature of the planet, $T_{eq}$, are depend on the Bond albedo of the planet $A$, the luminosity of the star $L_{star}$, the planet's semi major axis $D$, as well as the eccentricity $e$, of the orbit and in turn the average stellar irradiation at the planet's location. A more eccentric orbit increases the annually averaged irradiation proportional to $(1 - e^2)^{-1/2}$ (see William & Pollard 2002). In our atmospheric models the greenhouse effect is a factor in the overall albedo calculations and is not called out explicitly in this paper.

The inner edge of the HZ denotes the location where the entire water reservoir can be vaporized by runaway greenhouse conditions, followed by the photo-dissociation of water vapor and subsequent escape of free hydrogen into space. The outer boundary denotes the distance from the star where the maximum greenhouse effect fails to keep $CO_2$ from condensing permanently, leading to runaway glaciation (Kasting et al. 1993). Note that at the limits of the HZ, the Bond albedo of a habitable planet is fully determined by its atmospheric composition and depends on the spectral distribution of the stellar irradiation (Fig. 1). For a planet with a dense atmosphere, like Earth, $T_{eq}$ has to be below 270 K and above 175 K to be habitable (discussed in detail in Kaltenegger & Sasselov 2011 and references therein).

To simply estimate if a planet is potentially habitable (175K $\leq T_{eq} \leq$ 270K), one can use eq.(1) to approximate $T_{eq}$ for Earth-like planets around stars with $T_{eff}$ from 7200 K to 3200 K using the maximum albedo derived for different cloud coverage from Fig.1.

$$T_{eq}= ( (1-A) L_{star} / (4\beta D^2(1-e^2)^{1/2}) )^{1/4} \quad (1)$$

$\beta$ represents the fraction of the planetary surface that reradiates the absorbed flux. $\beta$ is 1 if the incident energy is uniformly reradiated by the entire surface of the planet, e.g. for a rapidly rotating planet with an atmosphere, like Earth (see results).

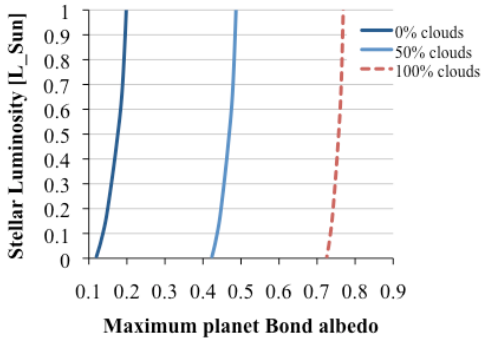

**Fig.1 Maximum planetary Bond albedo at the inner edge of the HZ derived for Earth-like atmosphere models for 0%, 50% and 100% clouds (from left to right).**

Fig.1 shows the maximum albedo derived from these models for an Earth-like atmosphere. The specific effects of clouds on the albedo depends on their height as well as particle size distribution (see e.g. Zsom et al. 2011, Goldblatt et al. 2011). The 100% cloud value (dashed lines) is used here to show the extreme effect of clouds on the HZ in accordance with work by Selsis (et al. 2007).

We model rocky planets with $H_2O/CO_2/N_2$ atmospheres, representative of geological active planets like Earth, to calculate the maximum Bond albedo as a function of irradiation and atmosphere composition and the edges of the HZ for HD 85512 b. These models represent rocky geological active planets and produce a dense $CO_2$ atmosphere at the outer edge, an Earth-like atmosphere in the middle, and a dense $H_2O$ atmospheres at the inner edge of the HZ.

Using eq. (1) and the maximum Bond albedo values from the atmosphere models, shown in Fig.1, provides the minimum $T_{eq}$ of the planet. Note that $T_{eq}$ does not correspond to any physical temperature at the surface or in the atmosphere. However it is very useful to assess habitability of a planet: if 175K $\leq T_{eq} \leq$ 270K, the planet could potentially be habitable (see e.g. Kaltenegger & Sasselov 2011). However a planet found in the HZ is not necessarily habitable, since many factors may prevent habitability like the lack of water or ingredients necessary for the emergence of life (see e.g. Zahnle et al. 2007, Selsis et al. 2007).

From the case studies by Kasting (Kasting et al. 1993) the inner $l_{in}$, and outer limit $l_{out}$, of the solar HZ can be extrapolated to stars with luminosity $L$, and effective temperature $T_{eff}$ between 3700 K and 7200 K using eq.(2):

$$l_x= (l_{x\_sun}-a_x T_{star}-b_x T_{star}^2) (L/L_{sun})^{1/2} \quad (2)$$

with $a_{in}$ = 2.7619 $10^{-5}$, $b_{in}$ = 3.8095 $10^{-9}$, $a_{out}$ = 1.3786 $10^{-4}$, $b_{out}$ = 1.4286 $10^{-9}$, and $T_{Star} = T_{eff} - 5700$, $l_x$ ($l_{in}$ and $l_{out}$) in AU, and $T_{eff}$ and $T_{star}$ in K.

Depending on the fractional cloud cover, the theoretical "water loss" limit

($T_{surf}$ = 373K) of the HZ of our Solar System assuming 0%, 50% and 100% cloud coverage, give $l_{in}$ as 0.95, 0.76 and 0.51 AU and $l_{out}$ as 1.67, 1.95 and 2.4 AU, respectively (see e.g. Selsis et al 2007). The inner limit for the 50% cloud case corresponds to the "Venus water loss limit", a limit that was empirically derived from Venus position in our Solar System (0.72 AU).

**Results**

Using the specifics of HD 85512 b we demonstrate the influence of the main parameters on its position in the HZ as well as the potential habitability of the planet. Fig. 2 relates the uncertainties in the measurements for RV planets, eccentricity, semi-major axis as well as stellar luminosity on its position in the HZ (right) and on the flux on top of the planetary atmosphere. $S_0$ is the solar flux received at the top of Earth's atmosphere at 1 AU, ($S_0$=1360 Wm$^{-2}$). The error on the y-axis shows the uncertainty in stellar luminosity and the error on the x-axis shows the uncertainty in the planet's position (left) as well as its irradiation (right). The limits of the HZ are shown here for 0%, 50% and 100% cloud coverage for comparison. Note that the x-axis in Fig. 2 (right) is increasing toward the left because the irradiation of the planet increases with decreasing distance from its host star and is not plotted on a log scale to allow detailed comparison of flux levels. The two panels in Fig. 2 relate the distance of the planet and the limits of the HZ to the flux received at the planet's position to compare it to Venus, Earth and Gl 581c.

The uncertainty of the derived eccentricity of 0.11 ± 0.1 is small and therefore does not changes the flux received at the planet's position significantly, only by +1.6% and -0.6%.

The uncertainty of the planet's semi major axis of 0.26 ± 0.005 AU change the incident flux on top of the planet's atmosphere by +4 % and –3.7 %. The biggest factor in the measurement uncertainties in stellar luminosity of 0.126 ± 0.008 that changes the flux at the planet's location by 6.3%.

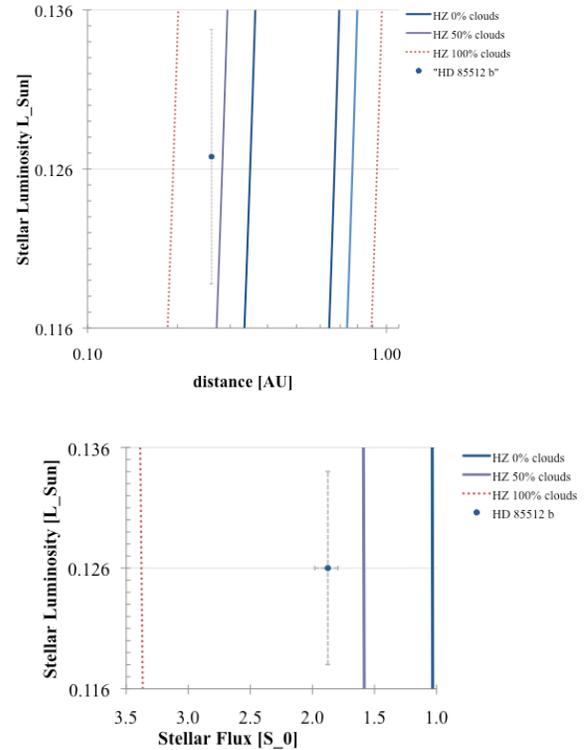

Fig. 2: Effect of the total measurement errors in semi-major axis and eccentricity and the resulting uncertainty (left) on the position in the HZ and (right) irradiation at the position of HD 58812 b in units of solar flux received at the top of Earth's atmosphere at 1 AU, $S_0$ =1360 Wm$^{-2}$. The error on the y-axis shows the uncertainty in stellar luminosity and the error on the x-axis shows the uncertainty in the planet's position (left) as well as its irradiation (right). The limits of the HZ are shown here for 0%, 50% and 100% cloud coverage as comparison.

The minimum flux on top of the planet's atmosphere is given for minimum stellar luminosity, minimum eccentricity, and maximum semi major axis. The maximum irradiation is given for maximum stellar luminosity, maximum eccentricity and minimum semi-major axis (shown in Fig. 2). These errors change the total incident radiation on the top of the planet's atmosphere by about +12 % and -10.7 %. Even the maximum increase in luminosity from uncertainties in the RV measurements does not put HD 85512 b outside the HZ.

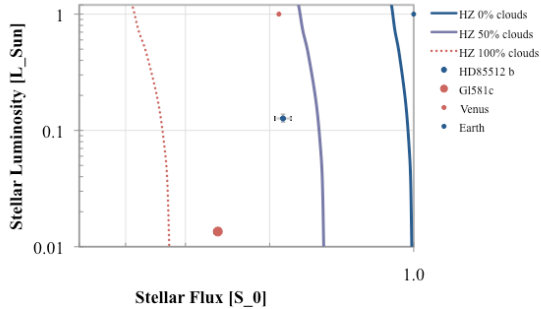

**Fig. 3 Stellar Luminosity in solar units, $L_{Sun}$, versus stellar flux on top of the planetary atmosphere, given in units of the flux received on the top of Earth's atmosphere, $S_0$. The extent of the inner edge of the HZ for the water loss limit for 0%, 50% and 100% (dashed line) cloud coverage (from right to left) and Earth, Venus and Gl 581c are added for comparison.**

Fig. 3 shows the stellar luminosity versus the stellar flux received on top of the planet's atmosphere in $S_0$. The error on the x-axis represents uncertainties in $e$ and the semi-major axis, while the error on the y-axis represent the uncertainty in stellar luminosity.

Earth, Venus as well as Gl 581c (Bonfield et al. 2005) are added for comparison. The limits of the HZ are also shown for 0%, 50% and 100% cloud coverage. Gl 581 c receives about 30% more flux than Venus, while HD 85512 b receives slightly less flux than Venus.

The total measurement uncertainties of +12 % and -10.7 % on the planet's irradiation change the overall flux received at HD 85512 b's position from 88% to 110% of the flux received by Venus.

Using the uncertainties on the stellar as well as planetary atmosphere model parameters, we calculate the planet's minimum equilibrium temperature shown in Fig.4.

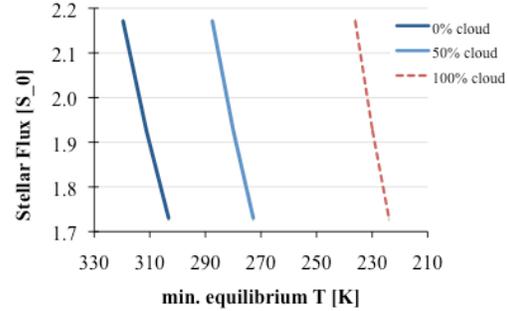

**Fig. 4: Minimum equilibrium temperature of HD 85512 b for 0%, 50% and 100% cloud coverage (left to right). The y axis corresponds to the total uncertainty of the stellar flux received at the planet's location +12%, -10.7% in $S_0$, due to measurement uncertainties in stellar luminosity as well as eccentricity and semi-major axis for HD 85512 b.**

Fig. 4 and Fig. 5 show the influence of the planetary atmosphere model parameters on habitability. In our atmospheric models the greenhouse effect is a factor in the overall albedo calculations and is not called out explicitly. Fig. 4 shows the influence of the atmosphere's maximum Bond albedo on $T_{eq}$, the curves represent models of 0%, 50% to 100% cloud cover respectively. We find that HD 85512 b could be potentially habitable if the planet exhibits more than 50% cloud coverage, assuming $\beta$=1.

There is no indication for a second planet in the system as of now and the eccentricity is expected to be zero, reducing the irradiation of the planet. A planetary albedo of 0.48 for a circular orbit and 0.52 +/- 0.05 for $e$ = 0.11 +/- 0.1 is needed to keep the equilibrium temperature below 270 K and the planet potentially habitable. As comparison Venus has a Bond albedo of about 0.75, while Earth and Mars have a Bond albedo of about 0.3, and 0.2 respectively. If clouds were increasing the albedo of HD 85512 b (for details see Zsom et al. 2011), its surface could remain cool enough to allow for liquid water if present.

Due to the close orbit of the planet to its star, tidal locking can be explored using the reradiation parameter β. As there is no indication for a second planet in the system for now, that could prevent the planet from being locked into synchronous rotation (see e.g. Leconte et al. 2010) or for high eccentricity, we explore the influence of the heat transfer by parameterizing $\beta$ to mimic the effect of potential tidal locking on the planetary climate. $\beta$ varies between 1 and 0.5, for a rapid rotating planet like the Earth to a planet that only reradiates the flux over half of its hemisphere respectively. Note that the atmosphere would have to be disconnected or non-existent for the extreme case of $\beta = 0.5$.

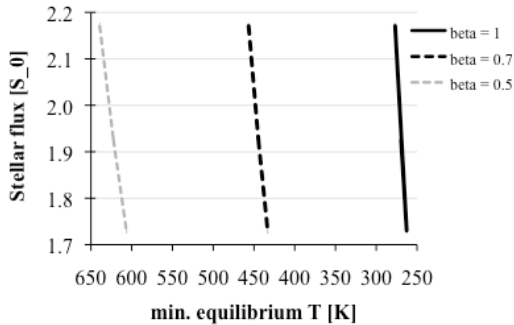

**Fig.5:** Effect of the heat transfer parameter β on the equilibrium temperature. The range of the stellar flux on the y-axis corresponds to the maximum and minimum irradiation received by the planet due to stellar and planetary parameter uncertainties for HD 85512 b.

Fig. 5 explores the influence of the heat redistribution parameter on the equilibrium temperature and in turn on habitability of a planet. It shows $T_{eq}$ for HD 85512 b as a function of $\beta$ factor for a Bond albedo of 0.52, what corresponds to the minimum albedo needed to keep $T_{eq}$ below 270K for $e$=0.11 and $\beta$=1. Parameterizing the heat transfer is only a rough approximation to show the resulting effect and detailed 3D models are needed to verify these results.

The reradiation factor has a strong influence on the amount of deposited flux as well as the resulting $T_{eq}$ of the planet's lit hemisphere. To keep $T_{eq}$ below 270K for $\beta = 0.7$, the planet would need a minimum Bond Albedo of 0.9. For lower values $T_{eq} > 270K$.

Note that the cloud fraction quoted in Fig. 1 needs to be maintained on the illuminated side of a synchronous rotating planet to achieve the Bond albedo.

For atmospheres like Earth's, it is unlikely that $\beta$ is much lower than 1. Detailed models have shown that even for planets in synchronous rotation direct illumination of only one hemisphere does not prevent habitability for planets with even modestly dense atmospheres like Earth (Haberle et al. 1996, Joshi et al. 1997, Joshi 2003, Edson et al. 2011, Wordsworth et al. 2011), provided atmospheric cycles transport heat from the dayside to the nightside.

The first planets below 10 Earth masses, with both mass estimates and radius measurements, have provided a wide range of observed radii and densities. Especially in the mass range below 5 Earth masses, two planets in the multiple planet system around Kepler 11 (see e.g. Borucki et al. 2011, Lissauer et al. 2011), Kepler 11 b and Kepler 11 f, with 4.3 and 2.3 Earth masses have radii of 1.97, 2.61 Earth radii and mean densities of 3.1, 0.7 g/cm$^3$ respectively. These derived densities allow substantial envelopes of light gases for this mass range. If HD 85512 b's density were low, like Kepler 11f, it would also be consistent with an extended H/He or

$H_2O$/H planet atmosphere. Recent atmosphere observations for a 6.55+/-0.98 Earth mass planet, GJ 1214 b, with a mean density of 1.8 g/cm$^3$ (Charbonneau et al 2010, Bean et al 2010, Desert et al. 2011), indicates either hazes or high cloud cover in an expanded atmosphere.

We do not discuss observed planets with a mean density comparable to Earth, e.g. Corot 7b (Leger et al. 2010) and Kepler 10b (Bathala et al. 2011), here, because of their close distance to their host star and resulting temperature range that is much higher than for HD 85512 b. Such temperature differences would most likely influence the atmospheric composition.

A larger sample will improve our understanding of this field and promises to explore a very interesting parameter space that indicates the potential co-existence of extended H/He and $H_2O$ dominated atmospheres as well as rocky planet atmospheres in the same mass and temperature range.

Advanced formation models aim to explain the existence of such a wide range of densities in this temperature mass range and are making rapid progress in recent years (see e.g. Mordasini et al. 2011, Ida & Lin 2010).

In this paper we only explore the potential habitability of HD 58812 b, assuming atmospheric models of rocky planets with $H_2O$/$CO_2$/$N_2$ atmospheres, like Earth, leaving the interesting question open whether a planet with an extended H/He/$H_2O$ could also be potentially habitable.

**Discussion**

Recent investigations of high precision radial velocity data samples have shown that between 20% and 50% of all sample stars exhibit RV variations indicating the presence of super-Earths or ice giants (Lovis et al. 2009, Howard et al. 2010, Mayor et al. 2011). Among them are two other possibly rocky planets around Gl 581 with masses of 5 and 8 $M_{Earth}$, respectively, lying at the edge of the HZ of their parent star (Udry et al. 2007, Mayor et al. 2009). We do not include the unfortunately unconfirmed planet Gl581g in our analysis (see also Forveille et al. 2011 and references therein).

The habitability of the most interesting planet until now, Gl 581 d, was discussed in detail by several groups (Selsis et al. 2007, Wordsworth et al. 2010, von Paris et al. 2010, Hu et al. 2011, Kaltenegger et al. 2011). Gl 581 d has a minimum mass of 7 $M_{Earth}$, a derived radius of 1.69 $R_{Earth}$ and orbits its star in 66.8 days on the outer edge of the HZ. Gl 581d requires several bar of $CO_2$ to remain habitable at that distance and irradiation. The $CO_2$ levels needed are consistent with a geologically active rocky planet model. Its minimum mass is about two times the minimum mass of HD 58812 b, increasing the probability of a substantial H/He envelope compared to HD 58812 b. As no radius is available for Gl 581d or HD 58812 b, the mean density of the planets can no be derived. To characterize the planet, a spectra of its atmosphere is needed, that could provide the distinction of a habitable world versus a Mini-Neptune (see e.g. DesMarais et al. 2002, Kaltenegger et al. 2010).

A hypothetical clear atmosphere model spectra for Gl 581 d in emission and transmission (Kaltenegger et al. 2011) was calculated to assess whether habitability could be detected remotely. The high $CO_2$ content in a habitable atmosphere model for Gl 581 d would reduces the signal of any observable biomarkers in its spectrum, but would in turn allow us to probe our concept of habitability and the corresponding $CO_2$ levels on the outer edge of the HZ.

Our atmosphere models for HD 58812 b only include the greenhouse effects of $H_2O$, $CO_2$ and $N_2$. Any additional greenhouse gas that could be present on such hot planets could in additional increase the temperature of the planets' surfaces. The effect of other greenhouse gasses on the inner edge of the HZ has not been explored yet and has therefore not been addressed here.

As discussed in the original discovery paper (Pepe et al 2011 for detailed discussion) the stellar parameters in table 1 may be affected by systematic errors, given the late spectral type of the star. Therefore two effective temperatures of 4715±102K and 4419K are obtained for this star for mass values of 0.73 and 0.65 $M_{sun}$ respectively. Both values are compatible with the error bars. The results shown in this paper are derived from the stars overall luminosity and are to first order not dependent on the exact temperature of the star, even though the maximum Bond albedo slightly changes (see Fig.1).

We assume that the planet's mass is similar to its minimum mass here, what leads to a surface gravity of about 1.4g for HD 58812 b, scaling its gravity with the mass and volume of a rocky planet. A higher mass and resulting higher gravity is expected to shift the inner edge of the HZ towards the star due to the gravity's influence on the water column and resulting IR opacity of the atmosphere, as well as on the lapse rate and Bond albedo. Atmosphere model calculations for a planet with surface gravity of 2.5g found that these effects roughly compensate and the inner edge of the HZ is only 3% closer to the host star (Kasting et al. 1993). For the expected gravity of HD 58812 b, this effect can be neglected.

## Conclusions

This paper outlines a simple approach to evaluate habitability of rocky planetary candidates from radial velocity searches by assuming models for rocky planets with $H_2O/CO_2/N_2$ atmospheres, like Earth.

We focus our analysis on HD 85512 b. We show the influence of the measurement uncertainties on its location in the Habitable Zone as well as its potential habitability. We find that HD 85512 b could be potentially habitable if the planet exhibits more than 50% cloud coverage. A planetary albedo of 0.48 +/- 0.05 for a circular orbit, and an albedo of 0.52 for e=0.11 is needed to keep the equilibrium temperature below 270K and the planet potentially habitable.

With its low mass and its incident irradiation slightly lower than Venus, HD 85512 b is, with Gl 581 d, the best candidate for habitability known to date. If clouds were increasing the albedo of HD 85512 b, its surface could remain cool enough to allow for liquid water if present. HD 85512 b is a planet on the edge of habitability.

## Acknowledgement

We are grateful to Dimitar Sasselov and Hans-Walter Rix for helpful comments and discussion. L.K. acknowledges support from NAI and DFG funding ENP Ka 3142/1-1.

**Table 1: Stellar and planetary parameter of HD 85512**

| Parameter [unit] | HD 85512 (Pepe et al. 2011) |
|---|---|
| Spectral Type | K5V |
| V | 7.67 |
| B − V | 1.156 |
| V variability | < 0.016 |
| Distance [pc] | 11.15 |
| $M_V$ | 7.43 |
| L [$L_{sun}$] | 0.126 ± 0.008 |
| [Fe/H] [dex] | −0.33 ± 0.03 |
| M [$M_{sun}$] | 0.69 |
| $T_{eff}$ [K] | 4715 ± 102 |
| log g [cgs] | 4.39 ± 0.28 |
| v sin i [km s$^{-1}$] | < 3 |
| $P_{rot}$ [days] | 47.13 ± 6.98 |
| age [Gyr] | 5.61 ± 0.61 |

| Parameter [unit] | HD 85512 b |
|---|---|
| P [days] | 58.43 (±0.13) |
| e | 0.11 (±0.10) |
| m*sin i* [$M_{Earth}$] | 3.6 (±0.5) |
| a [AU] | 0.26 (±0.005) |
| Teq [K] | 225 - 320 |